\documentclass[rnote]{aa}

\usepackage{verbatim}
\usepackage{graphicx}
\usepackage{epstopdf}
\usepackage{subfigure}
\usepackage{hyperref}
\usepackage{mathtools}

\title{The Coronal Temperatures of Low-Mass Main-Sequence Stars}

\titlerunning{The Coronal Temperatures of Low-Mass Main-Sequence Stars}

\author{C. P. Johnstone \and M. G\"{u}del}

\institute{
University of Vienna, Department of Astrophysics, T\"{u}rkenschanzstrasse 17, 1180 Vienna, Austria \label{vienna}
}

\abstract{}{
We study the \mbox{X-ray} emission of low-mass main-sequence stars to derive a reliable general scaling law between coronal temperature and the level of \mbox{X-ray} activity.
}{
We collect \emph{ROSAT} measurements of hardness ratios and \mbox{X-ray} luminosities for a large sample of stars to derive which stellar \mbox{X-ray} emission parameter is most closely correlated with coronal temperature. 
We calculate average coronal temperatures for a sample of 24 low-mass main-sequence stars with measured emission measure distributions (EMDs) collected from the literature.
These EMDs are based on high-resolution \mbox{X-ray} spectra measured by \emph{XMM-Newton} and \emph{Chandra}. 
}{
We confirm that there is one universal scaling relation between coronal average temperature and surface \mbox{X-ray} flux, $F_\text{X}$, that applies to all low-mass main-sequence stars.
We find that coronal temperature is related to $F_\text{X}$ by $\bar{T}_\text{cor} = 0.11 F_\text{X}^{0.26}$, where $\bar{T}_\text{cor}$ is in MK and $F_\text{X}$ is in \mbox{erg s$^{-1}$ cm$^{-2}$}.
}{}

\begin{document}

\maketitle

% ----------------------------------------------------------------------------------------------------------------------------------------------------------------------------------

\section{Introduction}

\begin{figure*}
\centering
\includegraphics[trim = 15mm 5mm 10mm 10mm, clip=true,width=0.32\textwidth]{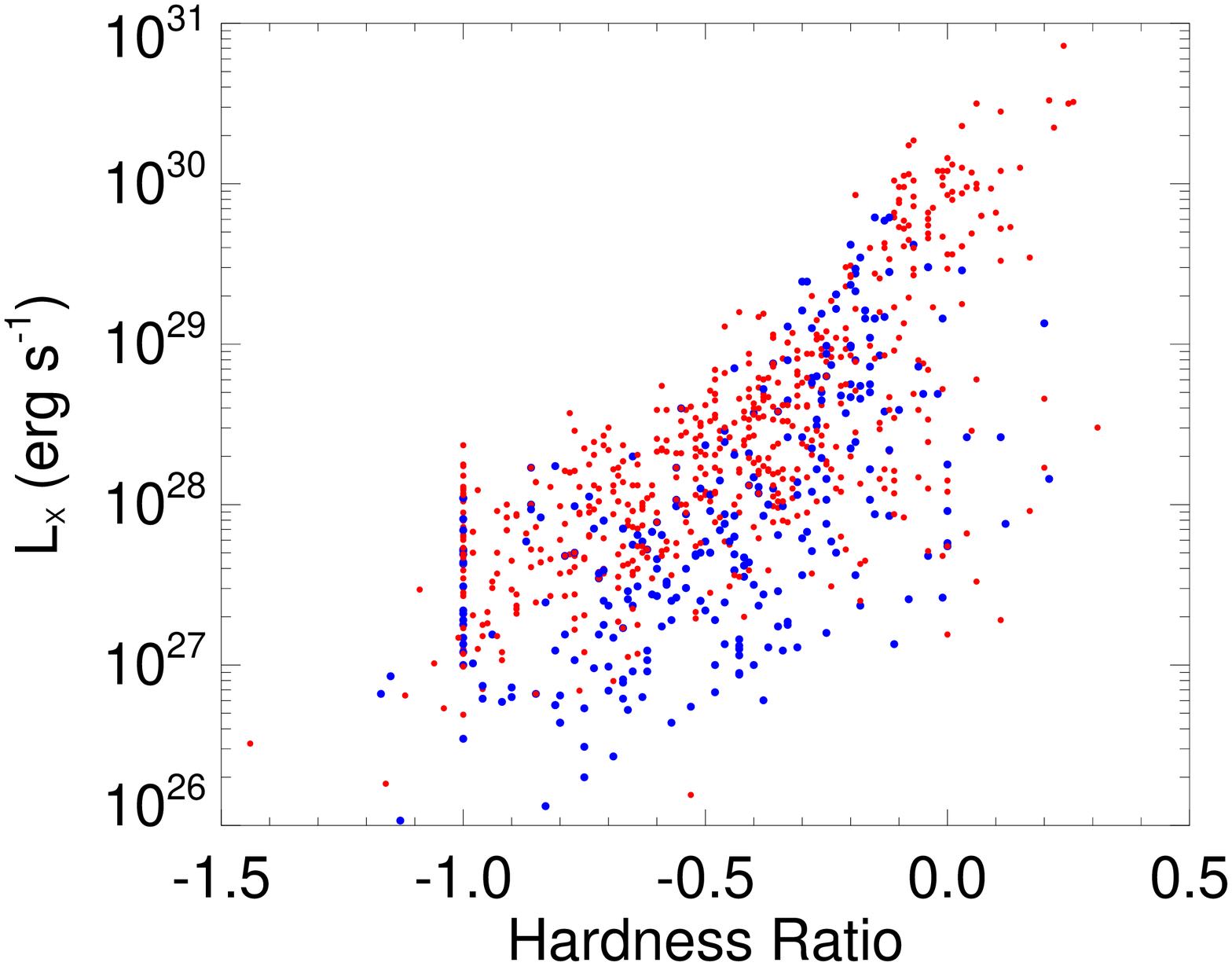}
\includegraphics[trim = 15mm 5mm 10mm 10mm, clip=true,width=0.32\textwidth]{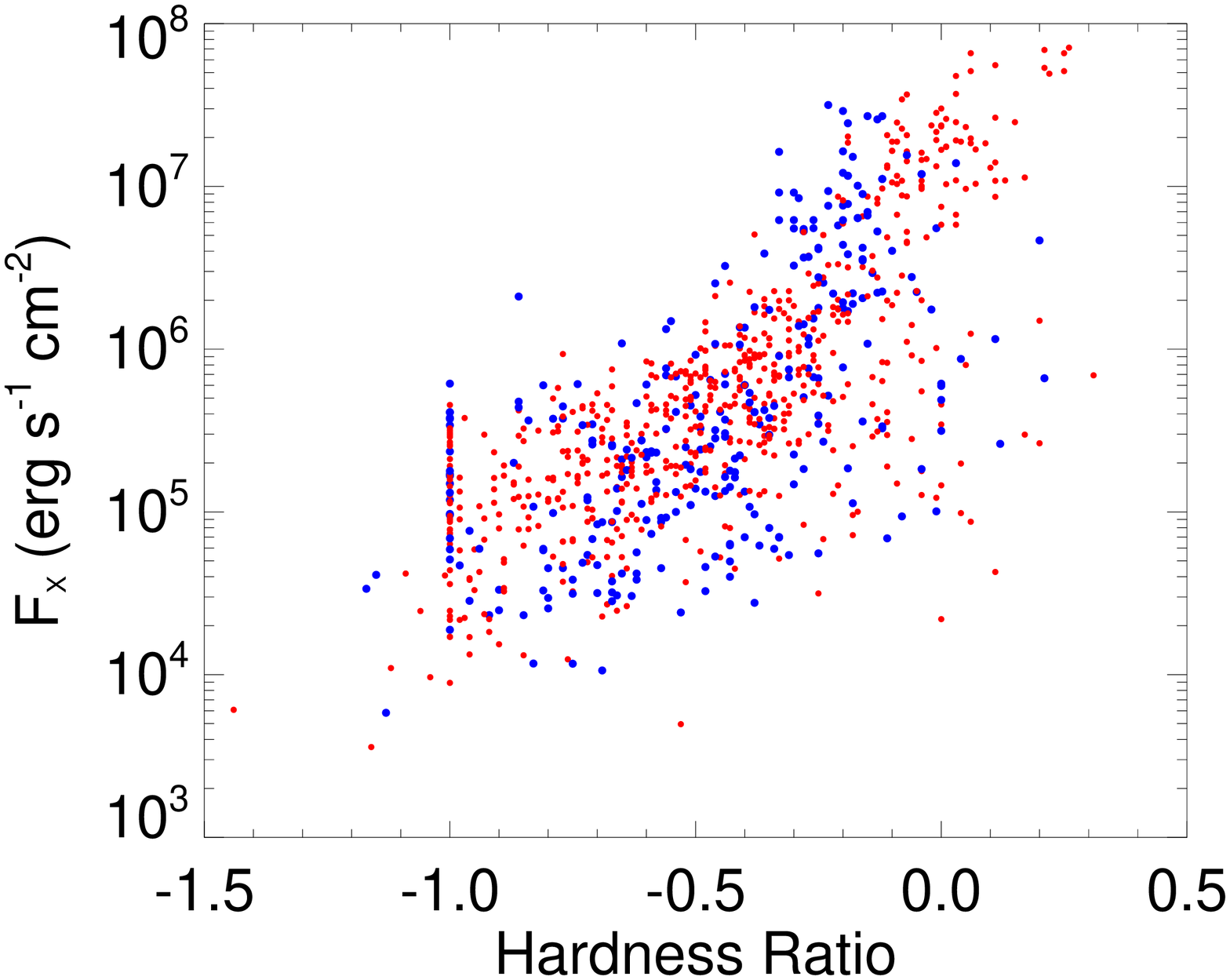}
\includegraphics[trim = 15mm 5mm 10mm 10mm, clip=true,width=0.32\textwidth]{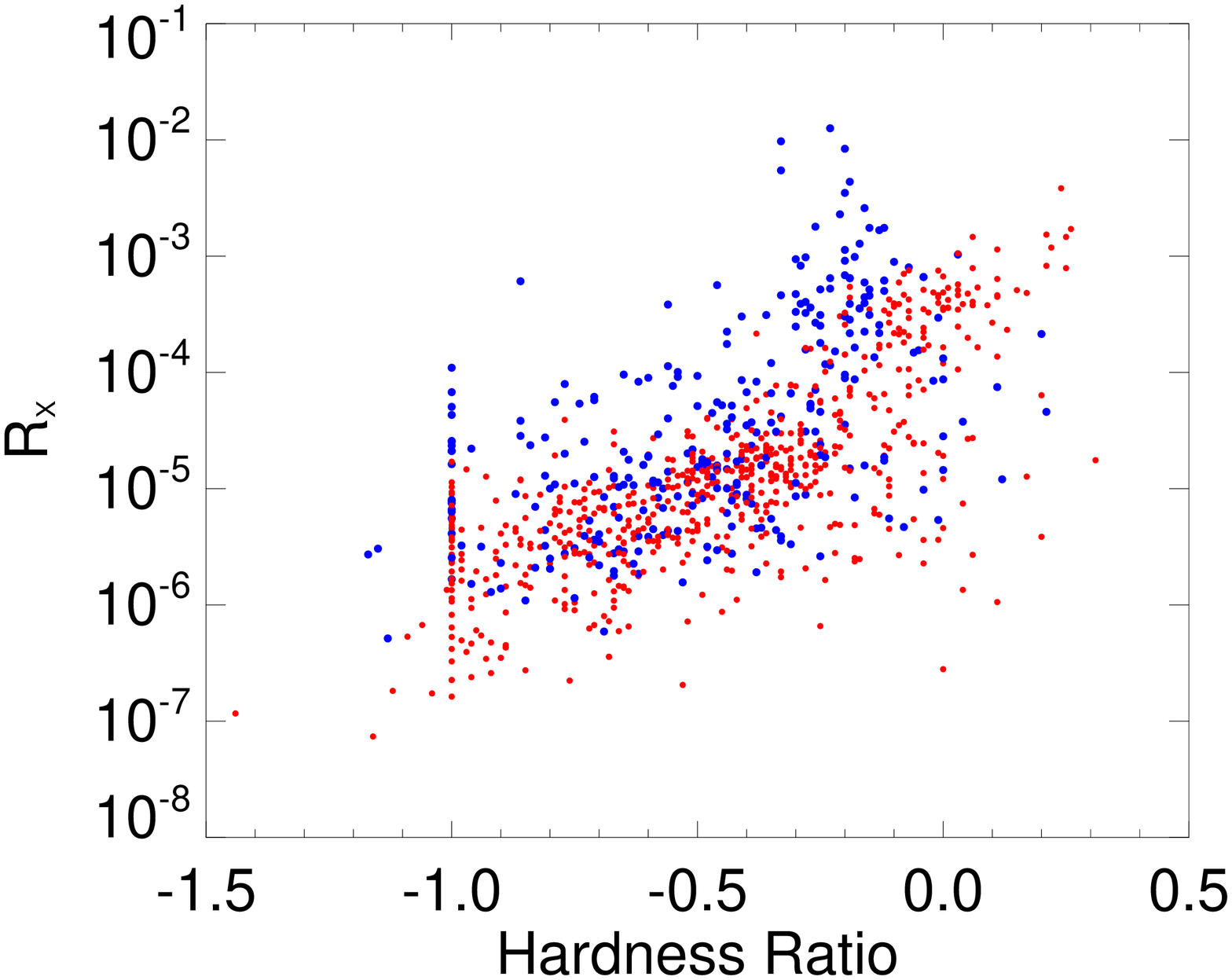}
\caption{
Plots showing how different measures of \mbox{X-ray} emission scale with hardness ratio for stars of different masses.
The hardness ratios and levels of \mbox{X-ray} emission were measured by \emph{ROSAT} and collected using the NEXXUS database (\citealt{2004A&A...417..651S}). 
The blue and red points represent stars with masses below and above 0.65~M$_\odot$ respectively. 
The three quantities of interest are \mbox{X-ray} luminosity (\emph{left panel}), \mbox{X-ray} surface flux (\emph{middle panel}), and \mbox{X-ray} luminosity divided by the bolometric luminosity (\emph{right panel}).
In order to calculate $F_\text{X}$ and $R_\text{X}$ we estimate the surface areas and the bolometric luminosities by assuming $R_\star \propto M_\star^{0.8}$ and $L_{\text{bol}} \propto M_\star^{3.9}$. 
}
\label{fig:XrayvsHR}
\end{figure*}

% some introductory remarks
The \mbox{X-ray} emission properties of low-mass main-sequence stars are strongly dependent on stellar mass and rotation (\citealt{1984A&A...133..117V}; \citealt{2003A&A...397..147P}).
Although the basic physical mechanisms responsible for heating coronae to \mbox{X-ray} emitting temperatures are not well understood, it is known empirically that coronal temperatures correlate well with levels of \mbox{X-ray} emission (\citealt{1983IAUS..102..165V}; \citealt{1984A&A...138..258S}; \citealt{1990ApJ...365..704S}; \citealt{1995ApJ...450..392S}; \citealt{1997ApJ...483..947G}; \citealt{1997A&A...318..215S}; \citealt{2005ApJ...622..653T}; \citealt{2007A&A...468..425T}). 
For example, \citet{2005ApJ...622..653T} analysed high-resolution \emph{XMM-Newton} spectra of six solar analogues with different levels of magnetic activity and showed that $L_\text{X} \propto \bar{T}_\text{cor}^{4.05 \pm 0.25}$, where $\bar{T}_\text{cor}$ is the emission measure weighted average coronal temperature (given by Eqn.~\ref{eqn:tempav}). 

An important question is which measure of \mbox{X-ray} emission correlates best with coronal temperature for stars with a range of surface areas and spectral types; the three obvious candidates are \mbox{X-ray} luminosity, $L_\text{X}$, \mbox{X-ray} luminosity normalised by the bolometric luminosity, $R_\text{X} = L_\text{X}/L_{\text{bol}}$, and \mbox{X-ray} surface flux, $F_\text{X} = L_\text{X} / (4 \pi R_\star^2)$.
\citet{1997A&A...318..215S} compared \emph{ROSAT} hardness ratio measurements for a sample of main-sequence F and G stars with similar measurements for a sample of main-sequence K and M stars and concluded that all low-mass main-sequence stars have the same relation between surface flux and spectral hardness. 
Similarly, \citet{1997A&A...320..525P} analysed \emph{ROSAT} spectra for a sample of stars and argued that $F_\text{X}$ is a better measure than $L_\text{X}$, though they did not consider the parameter $R_\text{X}$.

The answer to which of these parameters is best correlated with coronal temperature could contribute to our understanding of the magnetic processes responsible for heating coronal plasma to X-ray emitting temperatures and to our understanding of the coronal properties of saturated stars. 
For example, if $R_\text{X}$ is the relevant parameter, then this would indicate that temperature scales somehow with how far a star is below the saturation threshold in rotation, given that saturation happens at a single mass-independent value of $R_\text{X}$.
It would further imply that all saturated stars have similar coronal temperatures.
On the other hand, if either $L_\text{X}$ or $F_\text{X}$ are the relevant parameters, then it would indicate that among stars that lie above the saturation threshold, lower mass stars have cooler coronae than their higher mass counterparts.
The answer can also important for our understanding of stellar radiation in X-rays since stars with hotter corona will emit more photons at higher energies for a given total X-ray flux. 
This could be important for our understanding of the influences of stars on the upper atmospheres of planets, given that the influence that high-energy radiation has on a planet is highly wavelength dependent (\citealt{2005ApJ...621.1049T}; \citealt{2009A&A...496..863C}; \citealt{2015Icar..250..357C}).

%Furthermore, it could indicate that coronal heating follows an efficiency law, such that X-ray features on the stellar surface that are brighter per unit area are also hotter. 
%If this is the case, then distinguishing which of these two parameters is relevant could shed light on the influence of stellar surface area on X-ray emission.  
%Since stellar X-ray emission is mostly dominated by discrete coronal structures, the X-ray luminosity is determined by the number of significant emitting regions and the luminosities of the individual regions; it is not \emph{a~priori} obvious that the stellar surface area should influence the X-ray luminosity.
%Consider two stars with different surface areas but the same coronal temperature.
%If they both have the same value of $L_\text{X}$, then this would imply that surface area is not important in determining the number of active regions on the stellar surface.
%If they both have the same value of $F_\text{X}$, then this would imply that the number of active regions is instead proportional to the stellar surface area. 

% explain aim of paper
In this paper, we analyse coronal temperatures for a large sample of stars with different masses to show that $F_\text{X}$ is indeed the best indicator of coronal temperature and to derive a general scaling law that can be used to predict $\bar{T}_\text{cor}$ for all low-mass main-sequence stars.
In Section~\ref{sect:ROSATstuff}, we analyse \emph{ROSAT} measurements of hardness ratio for a sample of nearby \mbox{X-ray} emitting stars.
In Section~\ref{sect:highresstuff}, we calculate $\bar{T}_\text{cor}$ for a sample of stars based on high-resolution \emph{XMM-Newton} and \emph{Chandra} spectra and show how it correlated with $F_\text{X}$.

\begin{table}
\centering
\begin{tabular}{ccccc}
Name &		 	$R_\star$ & 	$F_\text{X}$ & 				$\bar{T}_\text{cor}$ & 	Reference \\
& 				(R$_\odot$) & 	(erg s$^{-1}$ cm$^{-2}$) &	(MK)  & \\
\hline
Solar Min & 		1.00 & 		$4.44 \times 10^3$ & 		0.97 & 				1\\
$\alpha$ Cen A  &	1.24 &     		$2.46 \times 10^4$   & 		1.49 & 				2\\
$\alpha$ Cen B &	0.84    &  		$5.60 \times 10^4$  &  		1.76 & 				2\\
Solar Max & 		1.00 & 		$7.73 \times 10^4$ & 		2.57 & 				1\\
SCR 1845 & 		0.10 & 		$2.30 \times 10^5$ & 		2.30 & 				3\\
$\beta$ Com & 		1.10$^{13}$ & 	$2.47 \times 10^5$ & 		3.89 & 				4\\
Prox Cen   & 		0.14$^{14}$ &	$3.36 \times 10^5$    &   		2.70 & 				5\\
$\xi$ Boo B    & 	0.61   &    		$4.13 \times 10^5$   &   		2.23 & 				6\\
70 Oph A & 		0.85 & 		$4.24 \times 10^5$ & 		3.39 & 				7\\
36 Oph B &   		0.59     &  		$4.31 \times 10^5$   &    		2.71 & 				7\\
36 Oph A & 		0.69 & 		$4.35 \times 10^5$ & 		2.82 &				7 \\
70 Oph B & 		0.66 & 		$4.65 \times 10^5$  & 		3.23 & 				7 \\
$\epsilon$ Eri & 	0.78 & 		$5.65 \times 10^5$ & 		3.48 & 				7 \\
$\pi^3$ Ori    &  	1.32$^{15}$&   	$8.61 \times 10^5$  &     		4.31 & 				8\\
$\chi^1$ Ori & 		1.02$^{13}$ &  	$1.41 \times 10^6$ & 		4.37 & 				4\\
$\kappa^1$ Ceti & 	0.93$^{13}$ & 	$1.70 \times 10^6$ & 		4.57 & 				4 \\
$\xi$ Boo A  &   	0.83 &      		$1.73 \times 10^6$   &    		4.37 & 				6\\
$\pi^1$ UMa &		 0.96$^{13}$ &	$2.05 \times 10^6$ & 		4.47 & 				4 \\
YZ CMi & 			0.36 & 		$2.73 \times 10^6$ & 		5.79 & 				9 \\
AD Leo   &   		0.37$^{16}$&   	$7.04 \times 10^6$    &  		6.39 & 				10\\
EV Lac     & 		0.36$^{17}$ &  	$7.44 \times 10^6$   &   		6.78 & 				10\\
AU Mic    & 		0.84$^{18}$ & 	$1.28 \times 10^7$   &    		7.09 & 				11\\
AB Dor     & 		1.10$^{19}$ & 	$1.39 \times 10^7$    &  		9.32 & 				12\\
EK Dra & 			0.91$^{13}$ & 	$2.39 \times 10^7$ & 		9.12 & 				4 \\
47 Cas B & 		1.00$^{13}$ & 	$4.04 \times 10^7$ & 		10.72 & 				4\\
\hline
\end{tabular}
\caption{
Properties of our sample of stars shown in Fig.~\ref{fig:telleschiscaling}.
From left to right, the columns correspond to the stellar radii that we use in our calculations, the $F_\text{X}$ values, the $\bar{T}_\text{cor}$ values, and the references for the studies from which these values were derived.
In most cases, we take the stellar radii from the studies that reported the X-ray luminosities and coronal temperatures; in cases where this was not possible, the superscripts on the $R_\star$ values give the references.
The references are as follows:
1.~\citet{2000ApJ...528..537P};
2.~\citet{2003A&A...400..671R}; 
3.~\citet{2010A&A...513A..12R}; 
4.~\citet{2005ApJ...622..653T}; 
5.~\citet{2004A&A...416..713G}; 
6.~\citet{2010ApJ...717.1279W}; 
7.~\citet{2006ApJ...643..444W}; 
8.~\citet{2013ApJ...768..122W}; 
9.~\citet{2007MNRAS.379.1075R}; 
10.~\citet{2005A&A...435.1073R}; 
11.~\citet{2003AdSpR..32.1149M}; 
12.~\citet{2013A&A...560A..69L};
13.~\citet{2007LRSP....4....3G};
14.~\citet{2009A&A...505..205D};
15.~\citet{2012ApJ...746..101B};
16.~\citet{2000A&A...354.1021F};
17.~\citet{2010ApJ...723.1558H};
18.~\citet{2009ApJ...698.1068P};
19.~\citet{2009A&ARv..17..251S}.
}
\label{tbl:sample}
\end{table}

\begin{figure}
\centering
\includegraphics[trim = 15mm 5mm 5mm 10mm, clip=true,width=0.49\textwidth]{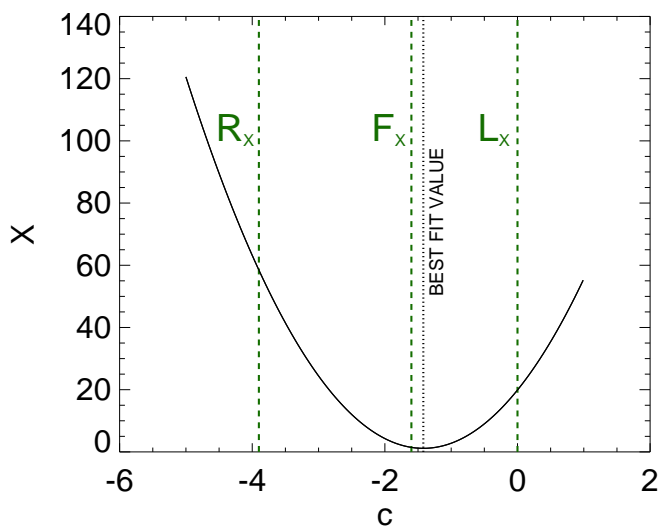}
\caption{
Best fit parameter, $X$, against the fit parameter $c$. 
For each value of $c$, we fit Eqn.~\ref{eqn:LxMstarTemp} to the sample of stars given in Table.~\ref{tbl:sample} and calculate $X$ from Eqn.~\ref{eqn:bestfitparam}.
The three vertical dashed lines show the value of $c$ that correspond to the different measure of X-ray emission and the dotted vertical line shows our best fit value of $c$.
}
 \label{fig:bestfit}
\end{figure}

\begin{figure*}
\centering
\includegraphics[trim = 15mm 0mm 5mm 10mm, clip=true,width=0.65\textwidth]{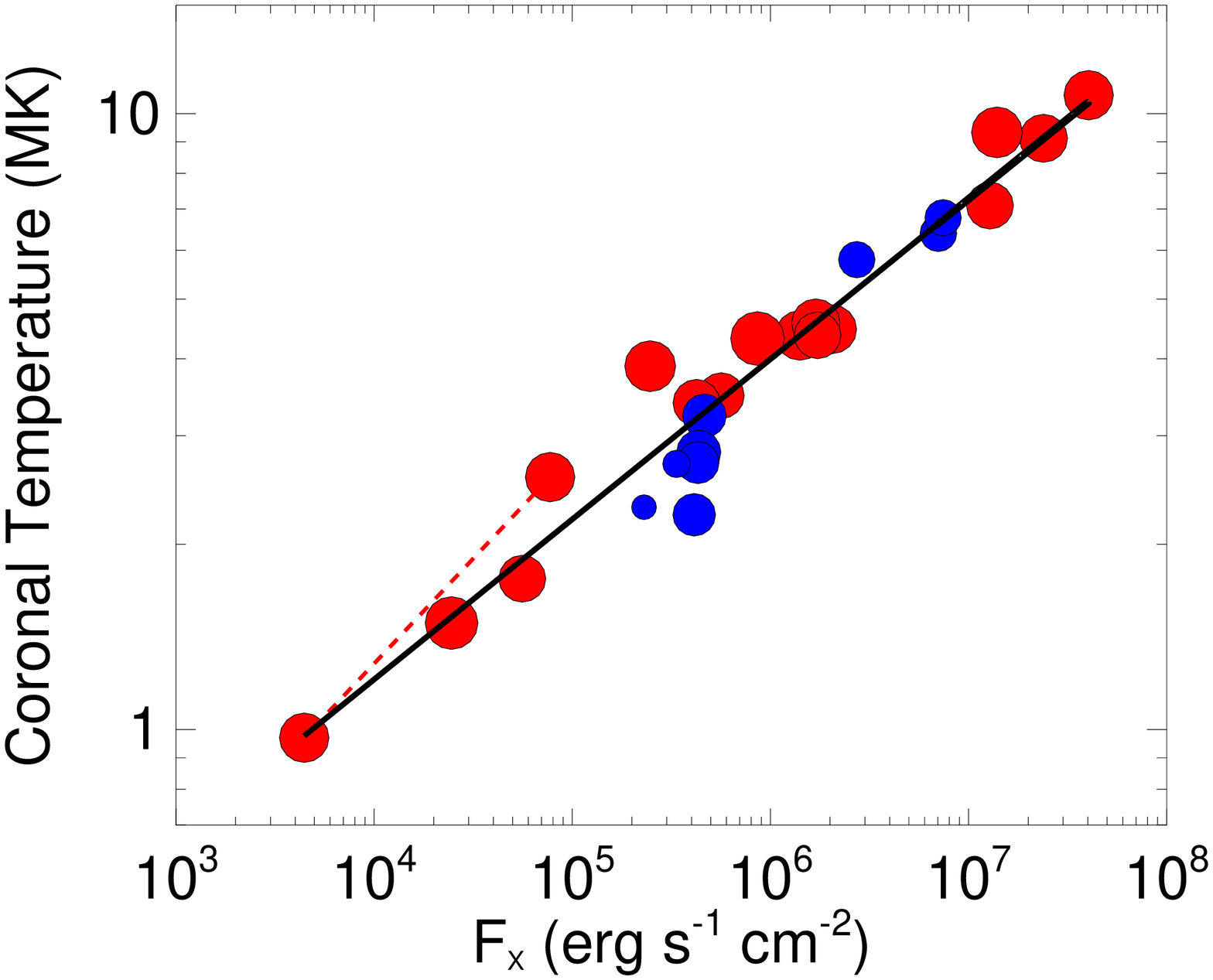}
\includegraphics[trim = 15mm 5mm 5mm 10mm, clip=true,width=0.4\textwidth]{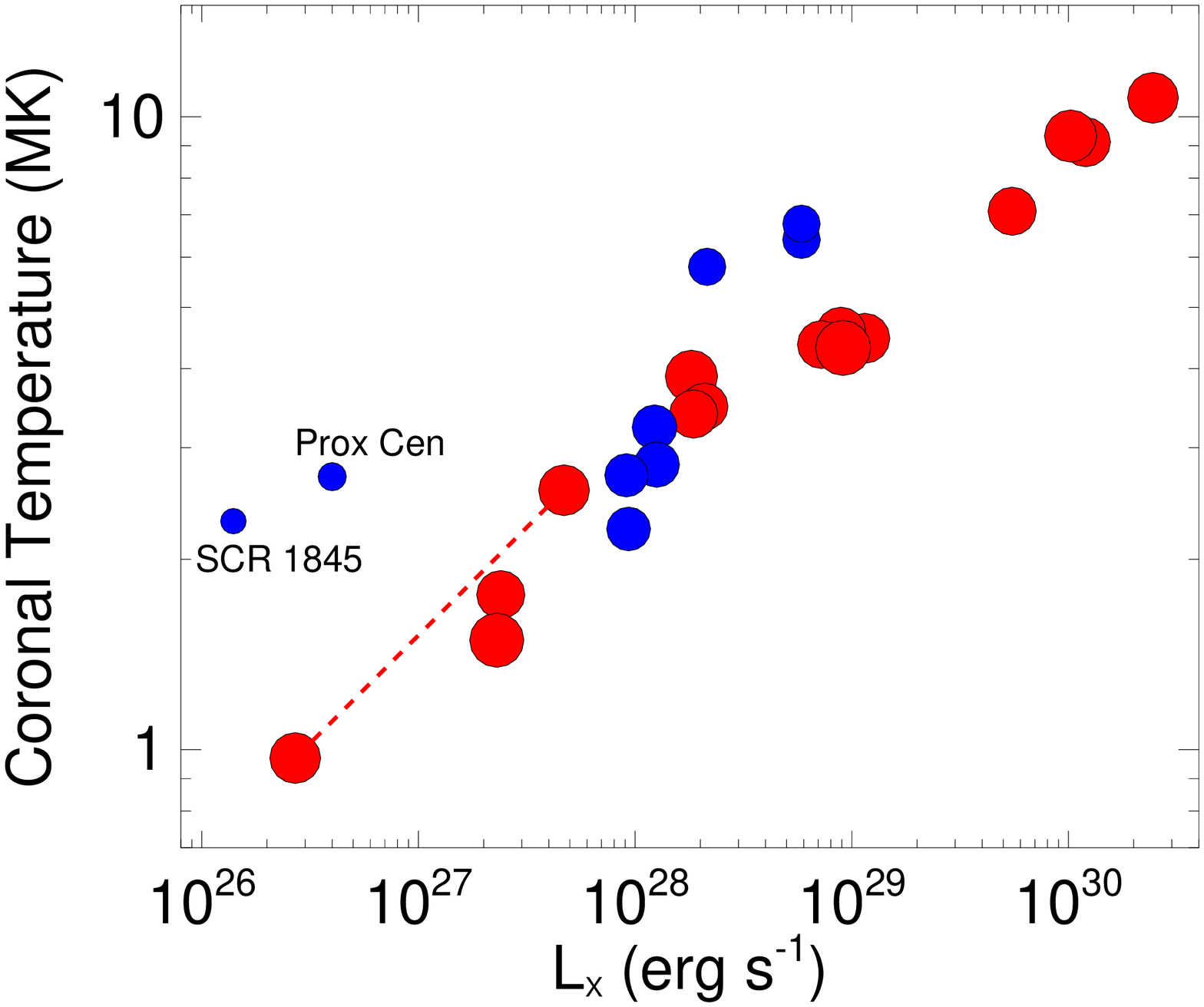}
\includegraphics[trim = 15mm 5mm 5mm 10mm, clip=true,width=0.4\textwidth]{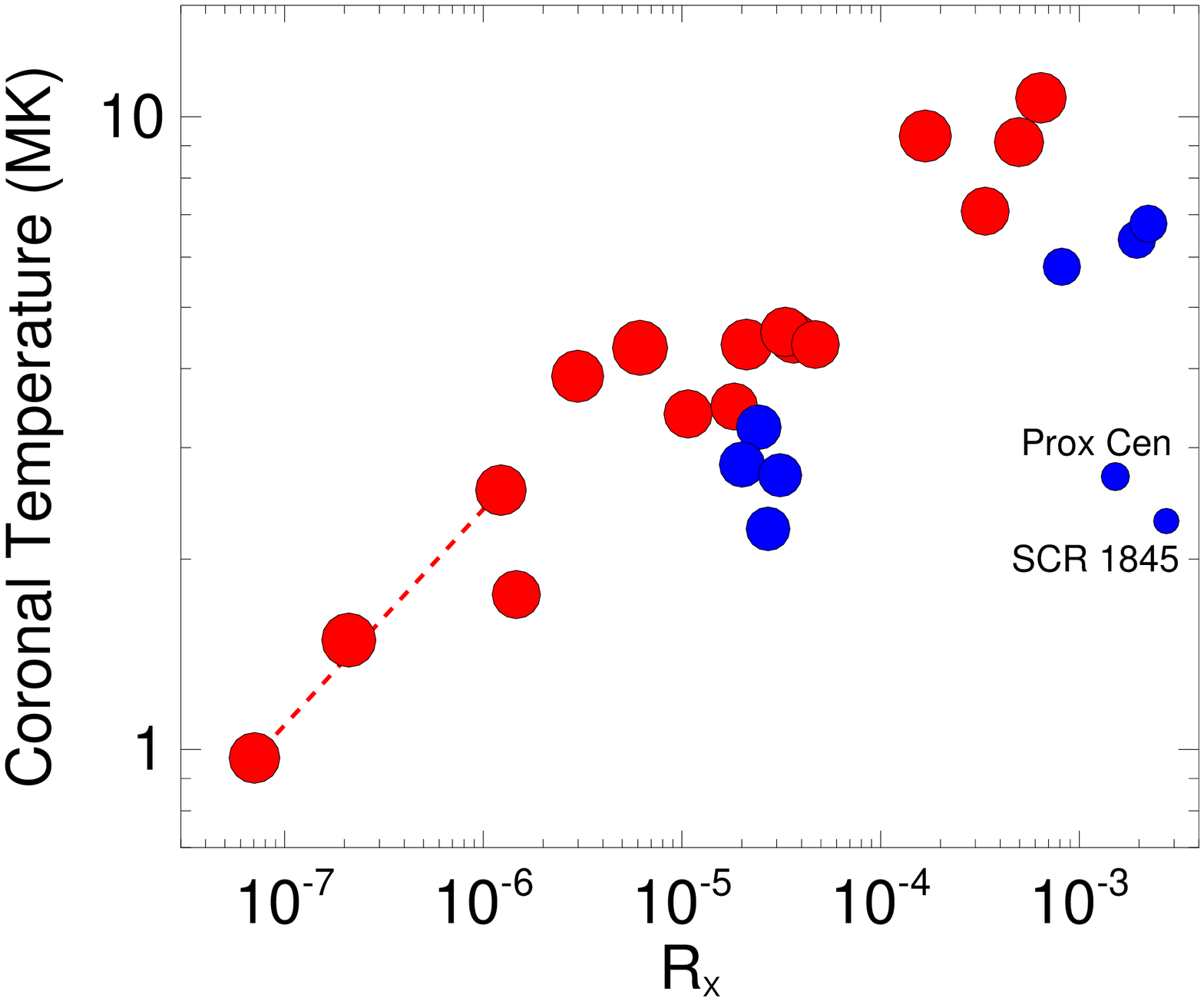}
\caption{
Average coronal temperature against $F_\text{X}$ (\emph{upper panel}), $L_\text{X}$ (\emph{lower-left panel}), and $R_\text{X}$ (\emph{lower-right panel}), for a sample of low-mass main-sequence stars.
As in Fig.~\ref{fig:XrayvsHR}, blue and red represent stars with masses below and above 0.65~M$_\odot$ respectively. 
The black line in the upper panel shows the relation \mbox{$\bar{T} \approx 0.11 F_\text{X}^{0.26}$}, where $\bar{T}$ is in MK and $F_\text{X}$ is in erg~s$^{-1}$. 
The two data points connected by the dashed line show the Sun at cycle minimum and cycle maximum.
The values and references for these stars are listed in Table~\ref{tbl:sample}.
}
 \label{fig:telleschiscaling}
\end{figure*}

\section{\emph{ROSAT} Hardness Ratio Measurements} \label{sect:ROSATstuff}

% what does coronal temperature scale with: Fx,Lx,Rx?

To explore the differences between $L_\text{X}$, $F_\text{X}$, and $R_\text{X}$ as indicators of coronal temperature, we collect a large sample of \emph{ROSAT} X-ray observations from the NEXXUS database (\citealt{2004A&A...417..651S})\footnotemark.
NEXXUS is a database of \mbox{X-ray} measurements for nearby stars and provides a comprehensive compilation of \emph{ROSAT} measurements, mostly from the \emph{ROSAT} All-sky Survey (RASS).
We collect X-ray luminosities and hardness ratios reported in the database (specifically the HR1 values) for all stars with masses between $\sim$0.2~M$_\odot$ and $\sim$1.1~M$_\odot$ with Position Sensitive Proportional Counter (PSPC) measurements (mostly as part of RASS).
Since hardness ratio is closely correlated with the temperature of the emitting plasma, they can be used to measure coronal temperature (\citealt{1995ApJ...450..392S}; \citealt{1997A&A...318..215S}).
The stellar masses are estimated by converting $(B-V)$ colours from the NEXXUS database using a conversion derived from the stellar evolution models of \citet{2007ApJ...671.1640A}.

The correlations between hardness ratio and the three measures of \mbox{X-ray} activity are shown in Fig.~\ref{fig:XrayvsHR}.
The red and blue points are for stars above and below 0.65~M$_\odot$ respectively (repeating this analysis with smaller mass bins leads to the same conclusion).  
Although there is overlap between the two mass bins in the $L_\text{X}$ and $R_\text{X}$ distributions, at a given temperature, higher mass stars are clearly likely to have higher values of $L_\text{X}$ and lower values of $R_\text{X}$ than lower mass stars.
On the other hand, the distributions for the two mass bins in the $F_\text{X}$ plot overlap excellently, providing a good visual indication that there is one mass-independent relation between $F_\text{X}$ and coronal temperature. 

\footnotetext{
The NEXXUS database can be found at \url{http://www.hs.uni-hamburg.de/DE/For/Gal/Xgroup/nexxus/index.html}.
}

\section{The Correlation Between \mbox{X-ray} Surface Flux and Coronal Temperature} \label{sect:highresstuff}

% our formula for scaling wind temperatures from Telleschi 
In this section, we show in a more comprehensive way that there exists one universal relation between coronal temperature and $F_\text{X}$ and derive a simple scaling law between the two quantities.  
For this purpose, we collect coronal temperature estimates from the literature based on multi-temperature emission measure distribution (EMD) fits to high-resolution \emph{Chandra} and \emph{XMM-Newton} spectra.
Our sample contains 24 stars in total with masses ranging from $\sim$0.1~M$_\odot$ to $\sim$1.2~M$_\odot$, including the Sun. 
For each EMD, we calculate the coronal average temperature by assuming that the logarithm of the average temperature is the emission measure weighted average of the logarithm of the temperatures of each component (\citealt{2007A&A...468..353G}; \citealt{2014MNRAS.437.3202J}), i.e.
\begin{equation} \label{eqn:tempav}
\log \bar{T}_{\text{cor}} = \frac{\sum_{i} \text{EM}_i \log T_i}{\text{EM}_\text{tot}},
\end{equation}
 where the sums are over all measured components of the emission measure distribution with each component having an emission measure of $\text{EM}_i$ and a temperature of $T_i$, and $\text{EM}_\text{tot} = \sum_i \text{EM}_i$. 
Stellar radii, $F_\text{X}$ values, $\bar{T}_{\text{cor}}$ values, and the references for our sample of stars are given in Table.~\ref{tbl:sample}.

We make the assumption that the coronal temperature, stellar mass, and $L_\text{X}$ can be related by
\begin{equation} \label{eqn:LxMstarTemp}
L_\text{X} M_\star^c = a\bar{T}_{\text{cor}}^b,
\end{equation}
where $a$, $b$, and $c$ are parameters to be fit to the sample.  
The interesting parameter here is $c$ since its best fit value is dependent on which measure of emission correlates best with coronal temperature.
If $L_\text{X}$ is the best parameter, then $c \approx 0.0$; if $F_\text{X}$ is the best parameter, then $c \approx -1.6$; if $R_\text{X}$ is the best parameter, then $c \approx -3.9$.
We define our goodness-of-fit parameter, $X$, for a given set of values for $a$, $b$, and $c$, as

\begin{equation} \label{eqn:bestfitparam}
X = \sum_i \gamma_i \left[    \log \left( L_{\text{X},i} M_{\star,i}^c\right) - \log(a) - b \log(T_i)    \right]^2 ,
\end{equation}
where the sum is over all stars given in Table.~\ref{tbl:sample} and $\gamma_i$ is a parameter that we use to weight the importance of certain stars in our fit.
In order to make the set of low mass stars have the same importance in our fit as the larger set of higher mass stars, we assume \mbox{$\gamma_i = 1$} for all stars with masses above 0.5~M$_\odot$ and \mbox{$\gamma_i = 4$} for all stars with masses below 0.5~M$_\odot$.
To find the value of $c$, we fit the free parameters in Eqn.~\ref{eqn:LxMstarTemp} for each value of $c$ between -5 and 1. 
In Fig.~\ref{fig:bestfit}, we show the best fit value of $X$ as a function of $c$. 
The best fit value of $c$ is -1.45, which shows that coronal temperature correlates with $F_\text{X}$ much better than with $L_\text{X}$ and $R_\text{X}$.%\footnotemark.
 
%\footnotetext{
%A similar analysis with the \emph{ROSAT} data presented in Section~\ref{sect:ROSATstuff} leads to a best fit at $c \approx -1$ which is more ambiguous. 
%Given that the EMD fits to high-resolution XMM-Newton and \emph{Chandra} spectra provide significantly more reliable measures of coronal temperature than \emph{ROSAT} hardness ratios, we do not consider this to be an uncertainty in our results. 
%} 
 
That $F_\text{X}$ is better than $L_\text{X}$ and $R_\text{X}$ can be easily seen in Fig.~\ref{fig:telleschiscaling} and is most clear from the two lowest mass stars in the sample, SCR~1845 and Proxima~Centauri.
Our best fit line in the upper panel of Fig.~\ref{fig:telleschiscaling} is given by
\begin{equation} \label{eqn:telleschi}
\bar{T}_{\text{cor}} \approx 0.11 F_\text{X}^{0.26},
\end{equation}
where $\bar{T}_{\text{cor}}$ is in MK and $F_\text{X}$ is in \mbox{erg s$^{-1}$ cm$^{-2}$}.
Our result is consistent with the relation provided by \citet{2005ApJ...622..653T} for just the solar analogues in the sample.
Despite the large range of masses, all of our stars fit this relation excellently, providing confirmation of our conclusion that one universal scaling law exists between $F_\text{X}$ and coronal temperature.

\section{Conclusions} \label{sect:conclusions}

It is very clear that the emission measure weighted average coronal temperature scales very closely with X-ray surface flux.
Our scaling law between $F_\text{X}$ and coronal temperature can be used to accurately estimate the coronal average temperature of any low-mass main-sequence star when the value of $F_\text{X}$ is known.
Alternatively, $F_\text{X}$ can be roughly estimated based on the star's mass and rotation rate (\citealt{2003A&A...397..147P}; \citealt{2011ApJ...743...48W}; \citealt{2014ApJ...794..144R}). 
The scaling law that we derive could be useful as input into coronal models (e.g. \citealt{2002MNRAS.333..339J}) and stellar wind models that scale wind temperature with coronal temperature (\citealt{2007A&A...463...11H}; \citealt{2015arXiv150306669J}; \citealt{2015arXiv150307494J}).

For a star with a given mass and radius, the level of \mbox{X-ray} emission is determined primarily by its rotation rate, with quickly rotating stars emitting more \mbox{X-rays} than slowly rotating stars until a certain threshold where the \mbox{X-ray} emission saturates (\citealt{1981ApJ...248..279P}; \citealt{1984A&A...133..117V}).
In the unsaturated regime, a star's X-ray luminosity scales well with its rotation period approximately as \mbox{$L_\text{X} \propto \Omega_\star^{2}$} (\citealt{2014ApJ...794..144R}), which combined with the assumption that \mbox{$R_\star \propto M_\star^{0.8}$}, implies that
\begin{equation}
\bar{T}_\text{cor} \propto M_\star^{-0.42} \Omega_\star^{0.52}.
\end{equation}
A similar result can be obtained using the scaling laws of \citet{2011ApJ...743...48W}.
Given that the saturation threshold appears to approximately be at a single mass-independent value of $R_\text{X}$, our result implies that in the saturated regime, low-mass stars have cooler coronae than high-mass stars.
Assuming \mbox{$L_\text{bol} \propto M_\star^{3.9}$} and \mbox{$R_\star \propto M_\star^{0.8}$} implies that among saturated stars
\begin{equation}
\bar{T}_\text{cor} \propto M_\star^{0.6}.
\end{equation}
Assuming a saturation Rossby number ($=P_\text{rot} / \tau_\text{c}$) of 0.13 (\citealt{2011ApJ...743...48W}) and a convective turnover time given by \mbox{$\tau_\text{c} = 15.49 \left( M_\star / M_\odot \right)^{-1.08}$~days} (\citealt{2014ApJ...794..144R}), the mass dependent satiation threshold is approximately given by 
\begin{equation}
\frac{\Omega_\star}{\Omega_\odot} = 13.53 \left( \frac{M_\star}{M_\odot} \right)^{1.08},
\end{equation} 
where we use $\Omega_\odot = 2.67 \times 10^{-6}$~rad~day$^{-1}$.

The mass dependence of coronal temperature in the saturated regime is an interesting consequence of the fact that coronal temperature depends on $F_\text{X}$ and not $R_\text{X}$ and can have consequences for the influence of stellar high-energy radiation on the upper atmospheres of planets.
Given that the location of the habitable zone is to first approximation determined by $L_\text{bol}$, all planets within the habitable zones around saturated stars should be exposed to approximately the same stellar X-ray and EUV fluxes regardless of the central star's mass.
However, the spectra of high-mass saturated stars is likely to be harder (i.e. more photons at higher energies) than the spectra of low-mass stars. 
The higher photon energies will mean that the radiation is likely to penetrate deeper into a planetary upper atmosphere, which could lead to differences in the atmospheric photochemistry and mass loss rates.

% ----------------------------------------------------------------------------------------------------------------------------------------------------------------------------------

\begin{acknowledgements}
The authors thank the referee for providing feedback on our research note and acknowledge the support of the FWF NFN project S11601-N16 ``Pathways to Habitability: From Disks to Active Stars, Planets and Life'', and the related FWF NFN subproject S11604-N16 ``Radiation \& Wind Evolution from T~Tauri Phase to ZAMS and Beyond''. 
This publication is supported by the Austrian Science Fund (FWF).
\end{acknowledgements}

% ----------------------------------------------------------------------------------------------------------------------------------------------------------------------------------

\bibliographystyle{aa}
\bibliography{mybib}

\end{document}